%% file: UEC.tex
\documentclass[
reprint,
pra,
superscriptaddress
]{revtex4-1}

\usepackage{graphicx,bm,array,longtable}

\newcommand{\rrscan}{r$^2$SCAN}
\newcommand{\sxc}{_\mathrm{xc}}
\newcommand{\mdf}{\mathrm{DFA}}
\newcommand{\mex}{\mathrm{exact}}
\newcommand{\br}{\bm{r}}

\begin{document}

\title{Unconventional Error Cancellation Explains the Success of Hartree-Fock Density Functional Theory for Barrier Heights}

\author{Bikash Kanungo}
\affiliation{Department of Mechanical Engineering, University of Michigan, Ann Arbor, MI 48109}

\author{Aaron D. Kaplan}
\affiliation{Lawrence Berkeley National Laboratory, Berkeley, CA 94720.}
\email{adkaplan@lbl.gov}

\author{Chandra Shahi}
\affiliation{Department of Physics and Engineering Physics, Tulane University, New Orleans, LA 70118}

\author{Vikram Gavini}
\affiliation{Department of Mechanical Engineering, University of Michigan, Ann Arbor, MI 48109}

\author{John P. Perdew}
\affiliation{Department of Physics and Engineering Physics, Tulane University, New Orleans, LA 70118}

\date{\today}

\begin{abstract}
    Energy barriers, which control the rates of chemical reactions, are seriously underestimated by computationally-efficient semi-local approximations for the exchange-correlation energy.
    The accuracy of a semi-local density functional approximation is strongly boosted for reaction barrier heights by evaluating that approximation non-self-consistently on Hartree-Fock electron densities, as known for about 30 years.
    The conventional explanation is that Hartree-Fock theory yields the more accurate density.
    This article presents a benchmark Kohn-Sham inversion of accurate coupled-cluster densities for the reaction H$_2$ + F $\rightarrow$ HHF $\rightarrow$ H + HF, and finds a strong, understandable cancellation between positive (excessively over-corrected) density-driven and large negative functional-driven errors (expected from stretched radical bonds in the transition state) within this Hartree-Fock density functional theory.
    This confirms earlier conclusions [Kaplan \textit{et al.}, J. Chem. Theory Comput. \textbf{19}, 532--543 (2023)] based on 76 barrier heights and three less reliable, but less expensive, fully-nonlocal density-functional proxies for the exact density.
\end{abstract}

\maketitle

Kohn-Sham density functional theory \cite{kohn1965} in principle yields exact ground-state energies and electron densities, while constraint-satisfying approximations to its exchange-correlation energy make useful predictions \cite{kaplan2023} over a vast materials space.
Understanding the successes and failures of such approximations is key to improving them.
It has been known for more than thirty years that the computationally efficient semi-local approximations, when implemented self-consistently, severely underestimate the barrier heights to gas-phase chemical reactions \cite{scuseria1992,oliphant1994,janesko2008,verma2012}, and that their accuracy for barriers is strongly boosted by performing a Hartree-Fock (HF) calculation and then replacing the HF exchange energy by the semi-local exchange-correlation energy evaluated on HF densities (and occupied orbitals if needed) \cite{scuseria1992,oliphant1994,janesko2008,verma2012}, a procedure known as ``Hartree-Fock density functional theory.''
More recently, this approach has been systematized as ``density-corrected density functional theory'' \cite{kim2013,wasserman2017,sim2018,vuckovic2019}, and has been shown to improve the average accuracy of other properties of main-group molecules \cite{santra2021} and to remarkably improve the binding energies of water clusters \cite{dasgupta2021,dasgupta2022}, when applied to constraint-satisfying semi-local functionals such as the Perdew-Burke-Ernzerhof generalized gradient approximation (PBE GGA) \cite{perdew1996} or the strongly-constrained and appropriately normed (SCAN) meta-GGA \cite{sun2015}.
The corresponding (non-self-consistent) Hartree-Fock density functionals are known as PBE@HF and SCAN@HF.
More generally, for any density functional approximation (DFA) there is a DFA@HF.
While Ref. \citenum{kim2013} rigorously defined density-driven error relative to the exact density, more recent work on density corrections has tended for practical reasons to take the HF density as a proxy for the exact density.

For many systems and properties, DFA@HF energy differences can be slightly more or slightly less accurate than those of self-consistent DFA@DFA.
For compact neutral atoms and molecules at equilibrium bond lengths (including the water monomer), there is graphical, statistical \cite{medvedev2017}, and energetic \cite{dasgupta2022} evidence that PBE and especially SCAN densities are modestly more accurate than HF densities.
SCAN exchange-correlation potentials for compact molecules are also reasonably accurate \cite{kanungo2021}.
But for large classes of systems and properties, DFA@HF energy differences are significantly and systematically more accurate than those of DFA@DFA.
For some of these systems and properties (dissociation limits of binding energy curves \cite{kim2015}, electron removal energies in small negative ions \cite{kim2011}), the reason is clearly that the more localized HF density yields the correct integer electron numbers on separated subsystems \cite{perdew1982} while the too-delocalized DFA density often yields spurious non-integer values.

The conventional explanation for large systematic improvements in energy differences from DFA@DFA to DFA@HF is that in these cases the self-interaction-free Hartree-Fock density is significantly more accurate than the self-consistent density of a semi-local approximation.
That explanation is indisputable for many cases, but is it also correct for barrier heights to chemical reactions and binding energies of water clusters, or is there a more correct explanation?

A forward barrier height is the energy difference between the transition state and the separated reactants, and a reverse barrier height is the energy difference between transition state and products.
The higher the barrier height, the slower the reaction.
The transition states of chemical reactions are typically stretched radicals.
The paradigm stretched radical is stretched H$_2^+$, where the semi-local functionals evaluated on the exact density can make the total energy severely too negative \cite{sun2016} for reasons discussed in Ref. \citenum{shahi2019}: the exact exchange-correlation hole is shared by two separated density fragments, while its semi-local approximation is not.
Thus the DFA error of the barrier height is not necessarily dominated by the error of the DFA density.
A more precise language is provided by the analysis of Burke, Sim, and collaborators \cite{kim2013,wasserman2017,sim2018,vuckovic2019}, who write the error of a self-consistent DFA for an energy or energy difference $E$,
\begin{equation}
  \Delta E_\mdf
  = E_\mathrm{DFA}[n_\mathrm{DFA}] - E_\mex[n_\mex]
  = \mathrm{FE} + \mathrm{DE},
  \label{eq:total_err}
\end{equation}
as the sum of a functional-driven error
\begin{equation}
  \mathrm{FE} = E_\mdf[n_\mex] - E_\mex[n_\mex]
  \label{eq:fe}
\end{equation}
and a density-driven error
\begin{equation}
  \mathrm{DE} = E_\mdf[n_\mdf]
    - E_\mdf[n_\mex].
  \label{eq:de}
\end{equation}
The exact electron density and exact total energy (but not the separate components of the total energy) are defined in the same way in density functional theory and in traditional quantum chemistry.
By the variational principle, DE is negative for a self-consistent DFA.
For a DFA@HF calculation, where $n_\mdf$ is replaced by $n_\mathrm{HF}$ in Eq. (\ref{eq:total_err}), we \emph{define} the analog of Eq. (\ref{eq:de}) by replacement of $n_\mdf$ by $n_\mathrm{HF}$,
\begin{equation}
  \mathrm{DE(DFA@HF)} = E_\mdf[n_\mathrm{HF}]
    - E_\mdf[n_\mex],
  \label{eq:de_at_hf}
\end{equation}
which can then be positive.
Equation (\ref{eq:fe}) remains unchanged by the same replacement, and the total error remains equal to FE $+$ DE.
With this replacement, Eq. (\ref{eq:de_at_hf}) is technically a ``density difference'' \cite{vuckovic2019} that vanishes when $n_\mathrm{HF} = n_\mex$, although it was called a ``density-driven error of $n_\mathrm{HF}$ in Ref. \citenum{nam2020}.
When DE(DFA@HF) is positive, the HF density over-corrects the DFA density; when $\mathrm{DE(DFA@HF)} \gg -\mathrm{DE(DFA)}$ it excessively over-corrects the DFA, and use of the HF density cannot be interpreted simply as a density correction to a DFA.

The precise evaluation of Eqs. (\ref{eq:fe})--(\ref{eq:de_at_hf}) would require not only the exact energy $E_\mex[n_\mex]$ and the exact density $n_\mex(\bm{r})$ (both well approximated in many cases by a coupled-cluster calculation), but also an inversion of the exact density to find the exact Kohn-Sham occupied orbitals for the evaluation of $E_\mdf[n_\mex]$ \cite{kanungo2021,nam2020}.
Accurate implementation of the last step is only practical for a limited number of systems, each of about 30 or fewer electrons. 
To better understand the errors of the 76 barrier heights in the BH76 test set \cite{zhao2005a,zhao2005b,goerigk2017}, Ref. \citenum{kaplan2023a} recently applied three fully-nonlocal proxies for the exact functional and density in Eqs. (\ref{eq:total_err})--(\ref{eq:de_at_hf}), chosen to satisfy two criteria: (1) accurate self-consistent barrier heights, and (2) nearly correct electron transfers due to nearly-linear variation of the total energy of a separated fragment between adjacent integer electron numbers \cite{perdew1982}.
(The semi-local approximations bend below the straight-line segment and are too de-localizing \cite{perdew1982,mori2006}, while Hartree-Fock bends above and is too localizing \cite{mori2006}.)
The proxy functionals were, in order of reliability, the long-range-corrected hybrid LC-$\omega$PBE \cite{vydrov2006}, a global hybrid of SCAN with 50\% exact exchange called SCAN50 or SX-0.5, and the self-interaction corrected SCAN-FLOSIC \cite{pederson2014}.
All three showed the same pattern: a large negative functional-driven error of PBE and SCAN, largely canceled by a large positive density-driven error when evaluated on the HF density.
The estimations of density-driven error (DE) in kcal/mol differed substantially between proxies, leaving some room for doubt.
For example, for the forward reaction in Table \ref{tab:BH} of this paper, they were (from Table S13 of Ref. \citenum{kaplan2023a}) $-1.3$ (PBE@LC-$\omega$PBE), $-4.9$ (PBE@SCAN50), $-6.4$ (PBE@SCAN-FLOSIC), although all were significantly different from $+11.3$ (PBE@HF) from Table \ref{tab:BH}, which uses an accurate CCSD(T) proxy.
The average over the three original proxies, -4.2, was not so different, from -2.2 (PBE) in Table \ref{tab:BH}.

Can we understand how all the BH76 transition states can have large negative functional-driven errors?
Such negative errors arise in the stretched radical H$_2^+$ (see Fig. 3 of Ref. \citenum{shahi2019}), while large positive functional-driven errors arise in the stretched, symmetry-unbroken singlet or non-radical H$_2$.
All of the BH76 transition states have stretched bonds, with total spins tabulated in Ref. \citenum{goerigk2017}.
Of 38 forward reactions, 23 involve an odd number of electrons, and their transition states are likely to be stretched radicals.
Of the remaining 15, 5 have non-singlet transition states that are also likely to be stretched radicals, and 10 have stretched singlet or non-radical transition states.
But none of these 10 dissociate to separated fragments with strong correlation between them.
6 of these 10 do not fragment in either the forward or reverse directions, and the remaining 4 have at most two fragments in either direction, at least one of which is closed-shell.
Thus none of the BH76 transition states appears to be like stretched H$_2$.

The work of Ref. \citenum{kaplan2023a} suggested that this unconventional error cancellation occurs strongly, widely and reliably for barrier heights, but the extent to which the proxies fairly represented the exact functional could still be questioned.
Here we will focus on the forward and reverse barrier heights of the BH76 reaction H$_2$ + F $\rightarrow$ HHF $\rightarrow$ H + HF, taking the coupled cluster CCSD(T)/aug-cc-pV5Z \cite{dunning1989} energies and densities \cite{scuseria1991} from the PySCF code \cite{sun2020} to be exact.
The resulting barrier heights differ by 0.2 kcal/mol or less from the W2-F12 ``exact'' values in BH76 \cite{goerigk2017}, which aim to reproduce CCSD(T) results in the complete basis-set limit \cite{karton2012}.
This work and Ref. \citenum{kaplan2023a} together permit a firm conclusion that, for many BH76 barrier heights, the Hartree-Fock density makes a density-driven error that largely cancels the substantial functional-driven error of PBE or SCAN.
This article also briefly discusses the possibility of a similar error cancellation in the water clusters, and presents a possible explanation for this unconventional error cancellation in molecules and molecular clusters.

With the help of the accurate coupled cluster method, we can evaluate the total DFA or DFA@HF error of a barrier height from Eq. (\ref{eq:total_err}).
But finding the separate functional-driven [Eq. (\ref{eq:fe})] and density-driven [Eq. (\ref{eq:de})] errors still requires an accurate determination of the Kohn-Sham orbitals that yield the CCSD(T) density, a challenging inverse problem.
For this, we use the method of Refs. \citenum{kanungo2021,kanungo2019}.
In this method, the inverse problem is formulated as a constrained optimization of the Kohn-Sham exchange-correlation potential $v_\mathrm{xc}(\bm{r})$ and solved using a convergent finite-element basis set.
Each finite element is a fifth-order Lagrange polynomial in the $x$, $y$, and $z$ directions.
For open-shell systems, we use a recent extension \cite{kanungo2023} of the inverse formulation with a spin-dependent exchange-correlation potential.
Self-consistent DFA and DFA@HF at the quadruple-zeta level can be found in Ref. \citenum{kaplan2023a}; we recompute these values at the quintuple-zeta level here.
All our density-functional calculations employ the separate up- and down-spin electron densities, not just the total density.
The DFA and DFA@HF calculations were treated as spin-unrestricted for F, H, and the HHF transition state; and as spin-restricted for H$_2$ and HF.
The local spin density approximation (LSDA) uses the parametrization of Ref. \citenum{perdew1992}.

Importantly, none of the functionals predicts a highly spin-contaminated transition state.
At the 5$\zeta$ level, $\langle \bm{S}^2 \rangle$ is 0.75 with the exact functional, 0.77 with HF, 0.75 with LSDA and PBE, and 0.76 with SCAN and \rrscan{}.

\begin{table}
  \centering
  \begin{tabular}{l|r|rr|r|rr} \hline
    & \multicolumn{3}{c|}{Forwards} & \multicolumn{3}{c}{Reverse} \\
   DFA & BH & FE & DE & BH & FE & DE \\ \hline
   LSDA & -23.7 & -20.7 & -4.4 & 25.4 & -3.8 & -4.7 \\
   LSDA@HF & -5.4 & -20.7 & 13.9 & 43.2 & -3.8 & 13.1 \\
   PBE & -12.6 & -11.8 & -2.2 & 24.8 & -6.8 & -2.3 \\
   PBE@HF & 0.9 & -11.8 & 11.3 & 37.6 & -6.8 & 10.5 \\
   SCAN & -7.4 & -7.8 & -1.0 & 22.0 & -10.6 & -1.2 \\
   SCAN@HF & 2.1 & -7.8 & 8.5 & 30.9 & -10.6 & 7.7 \\
   r$^2$SCAN & -6.9 & -7.3 & -1.0 & 23.8 & -8.9 & -1.3 \\
   r$^2$SCAN@HF & 2.5 & -7.3 & 8.5 & 32.6 & -8.9 & 7.6 \\  \hline
   CCSD(T) & 1.4 & 0.0 & 0.0 & 33.9 & 0.0 & 0.0 \\
  \hline
  \end{tabular}
  \caption{Barrier heights (BHs) and their functional-driven errors (FEs), and density-driven errors (DEs) for the reaction H$_2$ + F $\rightarrow$ HHF $\rightarrow$ H + HF.
  All units are kcal/mol.
  (1 Hartree $\approx 627.5$ kcal/mol; 1 eV $\approx 23.06$ kcal/mol.)
  FEs and DEs are computed by taking the CCSD(T)/aug-cc-pV5Z energies and densities as exact.
  The strong density sensitivity (absolute change of BH from LSDA to LSDA@HF $\gg$ 2 kcal/mol) is often taken as an indicator of the need for HF density correction \cite{vuckovic2019}.
  However, as $\mathrm{BH(DFA)} - \mathrm{BH(DFA@CCSD(T))}$ is about 1 kcal/mol for SCAN and \rrscan{} (see Table S1 of the Supplemental Materials), this should not be a highly density-sensitive system for the meta-GGAs.
  The sum of FE and DE yields the total error with reference to the CCSD(T)/aug-cc-pV5Z BH.
  }
  \label{tab:BH}
\end{table}

Table \ref{tab:BH} shows our numerical results.
The coupled cluster ``exact'' barrier heights are much smaller for the forward reaction than for the reverse.
The semi-local functionals severely underestimate the barrier heights, but there is overall improvement from LSDA to PBE to SCAN and its more computationally-efficient twin \rrscan{} \cite{furness2020}.
For these self-consistent DFAs, both FE of Eq. (\ref{eq:fe}) and DE of Eq. (\ref{eq:de}) are negative, but FE is typically much more negative.
From DFA to DFA@HF, the too-delocalized DFA density is replaced by the too-localized Hartree-Fock density, and DE becomes strongly positive, cancelling most of FE, especially for the more sophisticated SCAN and \rrscan{}.
This is the same error pattern found for the full BH76 set from the proxy-exact estimates of Ref. \citenum{kaplan2023a}.
By this energetic measure, the Hartree-Fock density for the transition state is actually much less accurate than the self-consistent DFA density.
But, as suggested at the end of Ref. \citenum{vuckovic2019}, there is in principle a DFA that yields the DFA@HF total energy and a self-consistent density expected to be more accurate than the HF density.

\begin{figure}
  \centering
  \includegraphics[width=\columnwidth]{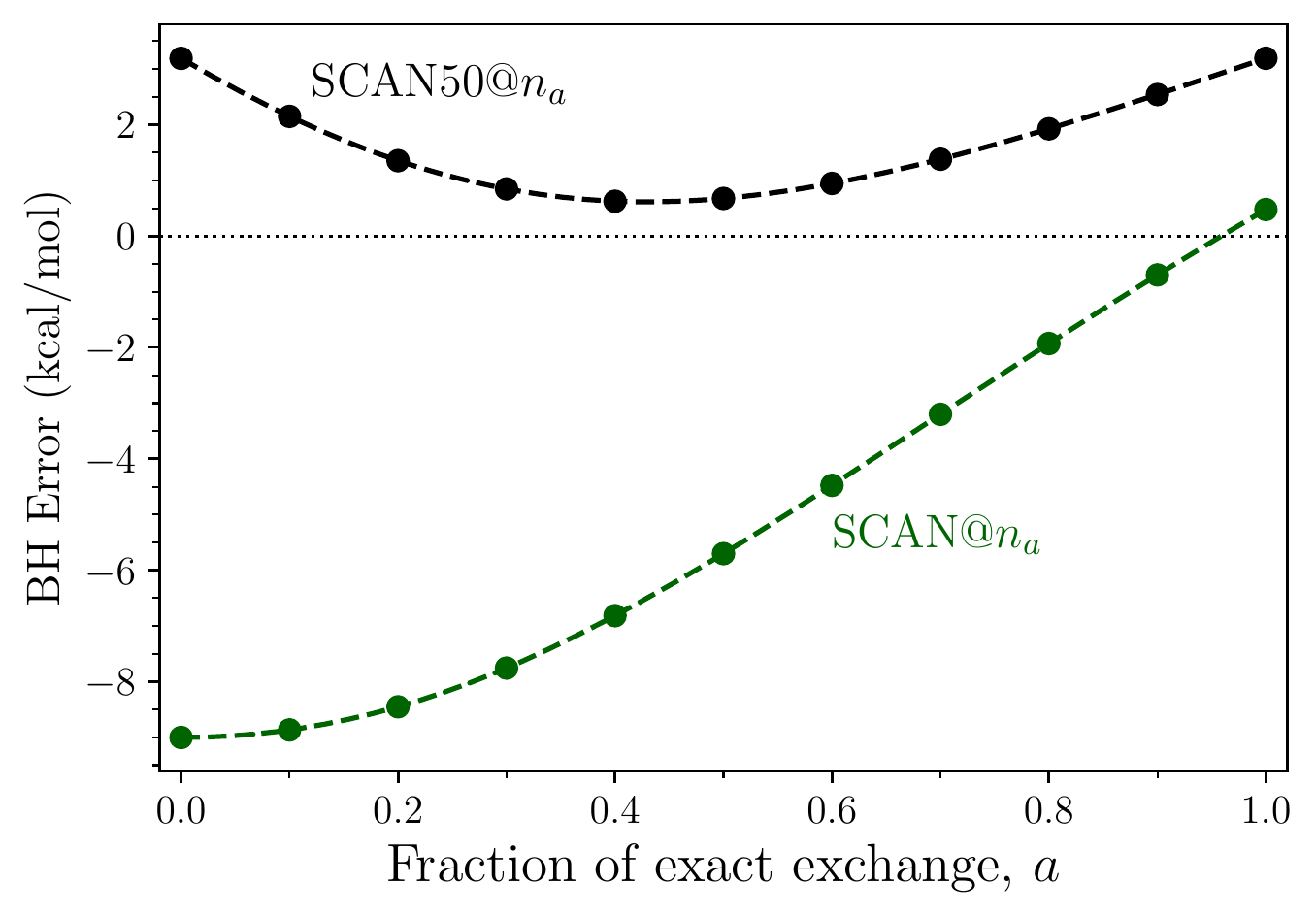}
  \caption{Error of the forward energy barrier height for the reaction H$_2$ + F $\rightarrow$ HHF $\rightarrow$ H + HF from SCAN (green) and proxy-exact SCAN50 (black), evaluated on a density $n_a$ that interpolates between the self-consistent SCAN density at $a=0$ and the HF density at $a=1$.
  That density is found self-consistently from the exchange-correlation functional of Eq. (\ref{eq:interp_xc}).
  }
  \label{fig:CT_ERR}
\end{figure}

Hartree-Fock DFT is a successful density correction to a DFA like SCAN when FE is small in magnitude and DE(DFA) is large, as in the dissociation limits of molecular binding energy curves or the electron affinities of atoms or small molecules, because in these cases the too-delocalized DFA density is qualitatively wrong while the too-localized HF density is qualitatively right.
In the barrier heights problem, however, DE(DFA) is much smaller in magnitude than FE, so that a true density correction would leave most of the total error uncorrected.
To understand what actually happens for the barrier heights, imagine a density $n_a$ computed self-consistently from a linear interpolation of the exchange-correlation energy
\begin{equation}
  E\sxc^\mdf[n_a] + a(E_\mathrm{x}^\mathrm{HF}[n_a] - E\sxc^\mdf[n_a])
  \quad (0 \leq a \leq 1).
  \label{eq:interp_xc}
\end{equation}
As Fig. \ref{fig:CT_ERR} shows, this is a small density variation around a minimizing density, for which $E_\mdf[n_a] \approx E_\mdf[n_\mdf] + C_\mdf a^2$, and $C_\mdf>0$.
The DFA ($a=0$) and HF ($a=1$) densities lie on a line in density space, on which localization increases with $a$, and the exact density lies roughly between them, at $a=q$ in the range from 0.3 to 0.5.
Then it is easy to see that
\begin{eqnarray}
  \mathrm{DE(DFA)} &\approx& - C_\mdf q^2, \\
  \mathrm{DE(DFA@HF)} &\approx& C_\mdf (1-q^2)
\end{eqnarray}
consistent with the signs and relative magnitudes of these two DEs in Table \ref{tab:BH}.

Why is the unconventional error cancellation between FE(DFA) and DE(DFA@HF) so good for barrier heights?
Such a reliable effect is unlikely to be accidental.
Figure \ref{fig:CT_ERR} also plots the proxy-exact SCAN50 evaluated on the density $n_a$.
The error of SCAN50 minimizes for $a \approx 0.43$ at a very small but positive error ($\approx 0.6$ kcal/mol), much as we would expect from the exact functional.
Taking SCAN50 to be a proxy for the exact functional's barrier height energy, the FE of SCAN, computed as the difference between the barrier-height errors in SCAN@$n_a$ and SCAN50@$n_a$, strongly decreases in magnitude as $a$ approaches 1, the HF limit.
The physical reason for this could be that SCAN and other semi-local functions become more accurate for a given density as that density becomes more localized and more HF-like.
Over the range $0 < a < 1$, SCAN varies much more strongly than proxy-exact SCAN50.

\begin{table} 
  \centering
  \begin{tabular}{l|r|rr|rr} \hline
    &  & \multicolumn{2}{c}{FE} & \multicolumn{2}{c}{DE} \\
   DFA & BE & CCSD(T) & r$^2$SCAN50 & CCSD(T) & r$^2$SCAN50 \\ \hline
   LSDA & -8.1 & -2.6 & -2.5 & -0.4 & -0.4 \\
   LSDA@HF & -6.9 & -2.6 & -2.5 & 0.7 & 0.7 \\
   PBE & -5.2 & 0.1 & 0.1 & -0.2 & -0.2 \\
   PBE@HF & -4.4 & 0.1 & 0.1 & 0.6 & 0.6 \\
   SCAN & -5.4 & -0.1 & -0.1 & -0.2 & -0.1 \\
   SCAN@HF & -4.7 & -0.1 & -0.1 & 0.4 & 0.5 \\
   r$^2$SCAN & -5.1 & 0.2 & 0.1 & -0.2 & -0.1 \\
   {\scriptsize r$^2$SCAN@HF} & -4.5 & 0.2 & 0.1 & 0.4 & 0.5 \\ \hline
   r$^2$SCAN50 & -4.8 &  &  &  &  \\
   CCSD(T) & -5.1 &  &  &  &  \\ \hline
  \end{tabular}
  \caption{Binding energies (BEs), functional-driven errors (FEs), and density-driven errors (DEs) for the water dimer, using the aug-cc-pVQZ basis set \cite{dunning1989}.
  In the CCSD(T) columns, FEs and DEs are computed by taking the CCSD(T) density to be $n_\mex$ in Eqs. (\ref{eq:fe})--(\ref{eq:de_at_hf}).
  In the r$^2$SCAN50 columns, FEs and DEs are computed using the self-consistent densities of the 50\% global hybrid of r$^2$SCAN, r$^2$SCAN50, as a proxy \cite{kaplan2023a} for the exact density $n_\mex$ in Eqs. (\ref{eq:fe})--(\ref{eq:de_at_hf}).
  In all cases, we take the self-consistent CCSD(T) binding energy to be $E_\mex[n_\mex]$.
  All values are in kcal/mol.
  }
  \label{tab:water_dimer}
\end{table}

Finally we turn to the (negative-definite) binding energy of a water cluster (H$_2$O)$_n$, defined as the energy of the bound cluster minus the energies of its $n$ separated H$_2$O monomers (at their optimized geometries).
SCAN is accurate for the relative energies of different hydrogen-bond networks, and even for the binding energy of the water dimer (H$_2$O)$_2$, but overestimates the binding of larger water clusters, reaching an error of about $-20$ kcal/mol for (H$_2$O)$_{20}$ clusters.
However, SCAN@HF reaches almost coupled-cluster accuracy for the binding energies of the larger water clusters \cite{dasgupta2021,dasgupta2022}.
Kohn-Sham inversion of a coupled cluster density for a large water cluster is computationally prohibitive at present, but we have done this for the water dimer in Table \ref{tab:water_dimer}.
While LSDA overbinds the water dimer by $-3$ kcal/mol, PBE, SCAN and especially \rrscan{} overbind by only a few tenths of a kcal/mol, in comparison to CCSD(T).
DFA@HF is more accurate than DFA for LSDA but not for PBE or SCAN.
Nevertheless, we still find that DFA@HF turns a small negative density-driven error of DFA into a substantially larger positive density-driven error.
In the larger water clusters, there might again be a cancellation in DFA@HF between negative FE and positive DE.
Table \ref{tab:water_dimer} also shows that the \rrscan{} 50\% global hybrid is a good proxy for the exact or CCSD(T) density, yielding almost the same FEs and DEs.
However, as its parent meta-GGA \rrscan{} makes essentially zero FE for the water dimer, admixture of exact exchange to correct errors in the \rrscan{} density introduces a more substantial FE to the \rrscan{}50 BE.
Composite methods like HF-\rrscan{}-DC4 \cite{song2023} (with a long-range dispersion correction) might be general-purpose practical solutions to this apparent catch-22.

To understand the density errors of DFA or DFA@HF,  $E_\mdf[n]$ must be used, as in Eqs. (\ref{eq:de}) and (\ref{eq:de_at_hf}), but there are many other ways to measure density errors that can lead to different conclusions about the relative accuracies of the DFA and HF densities.
For the neutral water dimer, Ref. \citenum{dasgupta2022} set up a plane perpendicular to the bond axis, such that a coupled cluster calculation put exactly 10 electrons on each side, and found electron transfer errors of opposite sign for semi-local DFAs and for HF.
Ref. \citenum{kaplan2023a} found the same behavior for several transition states.
The errors were small in magnitude, and smaller for HF than for a few DFAs.
In the cases studied here, $E_\mdf[n_\mathrm{HF}] - E_\mdf[n_\mex]$ is strongly positive, but that does not rule out $E_\mex[n_\mathrm{HF}] - E_\mex[n_\mdf]$ being negative; the HF density could be better than the DFA density in the sense of the exact density functional variational principle.
That said, Fig. S1 of the Supplemental Material shows an independent measure by which the density error of H...H...F  decreases from Hartree-Fock to SCAN to CCSD(T).

In summary, we have shown that DFA@HF works for the barrier heights to chemical reactions, and have suggested that it works for the binding energies of larger water clusters, not because the Hartree-Fock density is more accurate than the self-consistent DFA density but because the Hartree-Fock density creates a positive and excessive over-correction of the DFA density-driven error that cancels much of the negative functional-driven error.
The large functional-driven error for barrier heights was estimated first in Ref. \citenum{kaplan2023a}, and has been refined and confirmed here.
It is clear from Refs. \citenum{kim2013,wasserman2017,sim2018,vuckovic2019} and from Eqs. (\ref{eq:total_err})--(\ref{eq:de}) that, when the functional-driven error of a DFA is large and its density-driven error is small in comparison, a true density correction cannot lead to high accuracy.
Future work will employ proxy-exact functionals to test this hypothesis for larger water clusters.
Clearly, improved functionals will need the right amount of fully nonlocal density dependence, in both the exchange-correlation energy and the exchange-correlation potential.
Self-interaction corrections \cite{kaplan2023a,shukla2023} to DFAs, while needing improvement for some properties, appear for barrier heights to get the right answer for the right reason, by significantly reducing both functional- and density-driven errors.

The conclusions of Ref. \citenum{kaplan2023a} (cancellation of functional- and density-driven errors) for the BH76 barrier heights in Hartree-Fock density functional theory were based upon three (hybrid or self-interaction-corrected) proxies for the exact densities, and have been confirmed here for the H$_2$ + F $\rightarrow$ H + HF barriers, and their accurate Kohn-Sham inversions.
A confirmation for the full BH76 set, using a more efficient but perhaps less accurate approach (orbital optimized MP2), has been made recently in Ref. \citenum{hernandez2023}.
Reference \cite{singh2023} provides recent confirmation of our findings, with an interesting analysis of barrier-height errors with and without a self-interaction correction.
The higher accuracy of GGA, meta-GGA, and hybrid functional densities over the Hartree-Fock density was demonstrated for isolated atoms \cite{medvedev2017} and for the dipole moments of molecules at equilibrium \cite{hait2018}.

For the barrier heights to chemical reactions, as for the binding energies of equilibrium molecules, the density-driven errors of self-consistent DFA calculations are small, as the variational principle applied to Eq. (\ref{eq:de}) would suggest, but the functional-driven errors of the barrier heights are large in magnitude, as in Table \ref{tab:BH}.

The Supplemental Material presents checks on the accuracy of our calculations, total energies, and the employed molecular geometries, plus a figure showing the density errors of HF, SCAN, and CCSD(T) relative to Brueckner coupled-cluster doubles (CCD).

\begin{acknowledgments}
  B.K and V.G acknowledge support from the Department of Energy (DOE) grant DE-SC0022241.
  J.P.P. acknowledges the support of National Science Foundation grant DMR-1939528.
  J.P.P. and C.S. acknowledge support from DOE grant DE-SC0018331.
  A.D.K. acknowledges the support of a Temple University Presidential Fellowship and the DOE-BES Materials Project Program, Contract No. KC23MP.
  This research includes calculations carried out on HPC resources supported in part by the National Science Foundation through major research instrumentation grant number 1625061 and by the US Army Research Laboratory under contract number W911NF-16-2-0189.
  J.P.P. and A.D.K. acknowledge discussions with Kieron Burke (who suggested the numerical tests in Table S1).
  J.P.P. acknowledges discussions with Michael Medvedev, Saswata Dasgupta, and Filipp Furche (who suggested using the Brueckner CCD density as a reference in Fig. S1).
  C.S. thanks Qiming Sun for his help with the PySCF code.
  B.K., A.D.K., and C.S. contributed equally to this work. V.G. and J.P.P. contributed to the analysis and writing.
\end{acknowledgments}

\bibliographystyle{apsrev4-2}
\bibliography{UEC}

\input{UEC_SM}

\end{document}

%% file: UEC_SM.tex
\clearpage

\setcounter{page}{1}
\setcounter{section}{0}
\setcounter{table}{0}
\setcounter{figure}{0}

\renewcommand{\thepage}{S\arabic{page}}
\renewcommand{\thesection}{S\arabic{section}}
\renewcommand{\theequation}{S\arabic{equation}}
\renewcommand{\thetable}{S\arabic{table}}
\renewcommand{\thefigure}{S\arabic{figure}}

\onecolumngrid
\section*{Supplemental Material:\\
Unconventional Error Cancellation Explains the Success of Hartree-Fock Density Functional Theory for Barrier Heights}

\twocolumngrid
\tableofcontents
\onecolumngrid

\section{Barrier heights}

\input{./tables/KS_INV_QA.tex}

\clearpage
\input{./tables/BH_devs.tex}

\input{./tables/toten.tex}

\section{Complete basis set extrapolation}

To assess how significant the errors are in using quintuple-$\zeta$ values for the total energies, we must perform complete basis set (CBS) extrapolations.
To extrapolate Hartree-Fock (HF) or density functional approximation (DFA) total energies (\textit{not} energy differences), we use an exponential formula \cite{feller1993}
\begin{equation}
  E(\ell_\mathrm{max}) = E_\mathrm{CBS} + A \exp( -B \ell_\mathrm{max} ),
  \label{eq:cbs_exp}
\end{equation}
with $\ell_\mathrm{max}$ the maximum value of the orbital angular momentum quantum number contained in the aug-cc-pV$\left(\ell_\mathrm{max}\right)$Z basis set \cite{dunning1989}.
To extrapolate CCSD(T) correlation energies, defined for a given basis set as
\begin{equation}
  E_\mathrm{c}^\mathrm{CCSD(T)}\left(\ell_\mathrm{max}\right) \equiv
    E^\mathrm{CCSD(T)}\left(\ell_\mathrm{max}\right)
    - E^\mathrm{HF}\left(\ell_\mathrm{max}\right),
\end{equation}
we use the two-point formula \cite{helgaker1997}
\begin{equation}
  E_\mathrm{CBS} = \frac{ \left[ \ell_\mathrm{max}^{(1)} \right]^3
    E_\mathrm{c}^\mathrm{CCSD(T)}\left(\ell_\mathrm{max}^{(1)}\right) -
    \left[ \ell_\mathrm{max}^{(2)} \right]^3 E_\mathrm{c}^\mathrm{CCSD(T)}\left(\ell_\mathrm{max}^{(2)}\right)}
    {\left[ \ell_\mathrm{max}^{(1)} \right]^3 - \left[ \ell_\mathrm{max}^{(2)} \right]^3}.
  \label{eq:cw_extrap}
\end{equation}
Quintuple-$\zeta$ and CBS limit-extrapolated barrier heights are presented in Table \ref{tab:cbs_bh}.
Total energies extrapolated to the CBS limit are presented in Table \ref{tab:cbs_pars}.

\input{./tables/CBS_BH.tex}

\input{./tables/DF_pars.tex}

\subsection{Barrier height geometries}

Tables \ref{tab:xyz_bh_H2}, \ref{tab:xyz_bh_hf}, and \ref{tab:xyz_bh_RKT10} present the geometries of the molecules H$_2$, HF, and H...H...F used in this work for all calculations.
The atoms H and F were placed at the origin $x=y=z=0$ \AA{}.

\input{./tables/H2_geom.tex}
\input{./tables/hf_geom.tex}
\input{./tables/RKT10_geom.tex}

\section{Water Dimer}

\subsection{Water dimer geometries}

\input{./tables/H2O_geom.tex}
\input{./tables/H2O_2_geom.tex}


\section{Discussion of single- and multi-reference methods, Hartree-Fock-orbital and Brueckner-orbital coupled cluster methods}

Full single-reference configuration interaction (CI) (which includes all possible excitations out of a single ground-state Slater determinant) is exact, because the set of all such excited Slater determinants is complete \cite{szabo1982}.
The single-determinant reference state is usually constructed from Hartree-Fock orbitals in CI and coupled-cluster methods, although other choices are possible.
The presence of low-lying excited states near the highest-occupied state can make a single-determinant reference state a poor starting point for describing the system.
One would then need a multi-reference (MR) starting point for the correlated wavefunction calculation. 
Thus multi-reference methods may be needed when the excitations considered by an approximate method are restricted to a limited set, such as single and double excitations for a system with more than two electrons.

The role of low-lying near-degenerate states has been observed \cite{li2006} for the strongly-correlated molecles BN and C$_2$, which have singlet-triplet splittings of about 0.4 and 2.0 kcal/mol, respectively.
In both cases, single-reference CCSD(T) reduces the \emph{absolute} errors of single-reference CCSD, but MR CCSD(T) was needed to obtain the correct energy ordering of the singlet and triplet in BN, and lower absolute errors in their splitting.

A few metrics have been proposed to determine (with some confidence) when a single-determinant reference is sufficient for treating electronic correlation.
Reference \cite{duan2020} discusses the relevant ranges of these metrics, which are commonly constructed from the CCSD ``amplitudes,'' which are essentially the expansion coefficients for different excited Slater determinants.
Two metrics, $T_1$ and $D_1$, require the single-excitation amplitudes, whereas $D_2$ requires the two-electron excitation amplitudes.
Then for a single-determinant reference to be sufficient, $T_1 < 0.02$, $D_1 < 0.05$, and $D_2 < 0.18$ for small organic molecules, such as the transition state H...H...F considered here.

We found $D_1(\mathrm{CCSD})=0.12$ for the H...H...F transition state, and 0.02 for H$_2$O.
This could suggest that a single Slater determinant is insufficient for H...H...F, however it is possible that single-reference CCSD(T) is sufficient here, as it recovers a greater fraction of electronic correlation than CCSD alone.
This would be consistent with the use of CCSD(T)-equivalent barrier heights as reference exact values in the BH76 subset of the GMTKN55 database \cite{goerigk2017}.
The reference values used there are from the W2-F12 method \cite{karton2012}, which aims to reproduce CCSD(T) results in the complete basis-set limit.
For the forward and reverse barrier heights with the H...H...F transition state, the differences between CCSD(T)/aug-cc-pV5Z and W2-F12 are about 0.2 kcal/mol.

References \cite{duan2020} and \cite{duan2022} use the CCSD(T) energy as a reference exact energy for systems that need MR treatment at the CCSD level.
That may be acceptable, unless the need for MR treatment at the CCSD level is very strong.
They also use
\begin{equation}
  \%E_\mathrm{corr}[\mathrm{(T)}] = 100\% \times
  \frac{E_\mathrm{c}^\mathrm{CCSD}}{E_\mathrm{c}^\mathrm{CCSD(T)} }
\end{equation}
as a measure of the need for MR treatment at the CCSD level (with more need when this measure is smaller).
For the H...H...F transition state, we have found this diagnostic to be 98\%, which does not suggest a strong need for a MR starting point.
Compare this with Fig. 1 of Ref. \cite{duan2022}, which observed that:
\begin{enumerate}
  \item The need for a MR starting point \emph{increased} as the strength of a chemical bond \emph{increased};
  \item The need for a MR starting point \emph{decreased} as the number of unpaired electrons \emph{increased};
  \item The need for a MR starting point \emph{decreased} as the bond length in a metal-He complex \emph{increased}.
\end{enumerate}
The second and third observations suggest that the MR character of H...H...F should not be too strong.

To determine the bond character in the transition state, we use the turning surface analysis of Ref. \cite{ospadov2018}.
Their metric uses the separation between atoms A and B in a molecule and the turning surface radii (of the exact Kohn-Sham potential) of isolated atoms A and B to determine the bond character between A and B.
Let $\br_\mathrm{A}$ and $\br_\mathrm{B}$ be their positions in the molecule, and $\tau_\mathrm{A}$ and $\tau_\mathrm{B}$ be the turning surface radii of the isolated atoms.
Then
\begin{equation}
  \beta(\mathrm{AB}) = \frac{|\br_\mathrm{A} - \br_\mathrm{B}|}
    {\tau_\mathrm{A} + \tau_\mathrm{B}}
\end{equation}
should be (approximately) less than 0.7 for a covalent bond; between 0.7 and 0.9 for a typical hydrogen bond; between 0.9 and 1.3 for an ionic bond, and greater than 1.3 for a dispersion bond \cite{ospadov2018}.
In the H...H...F transition state: $\beta(\mathrm{H}_1 \, \mathrm{H}_2) = 0.37$, or strongly covalent bonding;
$\beta(\mathrm{H}_1 \, \mathrm{F}) = 0.72$, or weak covalent bonding; and $\beta(\mathrm{H}_2 \, \mathrm{F}) = 1.07$, likely a weak hydrogen bond.
As the bonding in the transition state is a mix of relatively weak bonds, we find all three criteria for adequacy of the single-determinant reference to be sufficiently met.

\begin{figure}
    \centering
    \includegraphics[width=0.6\columnwidth]{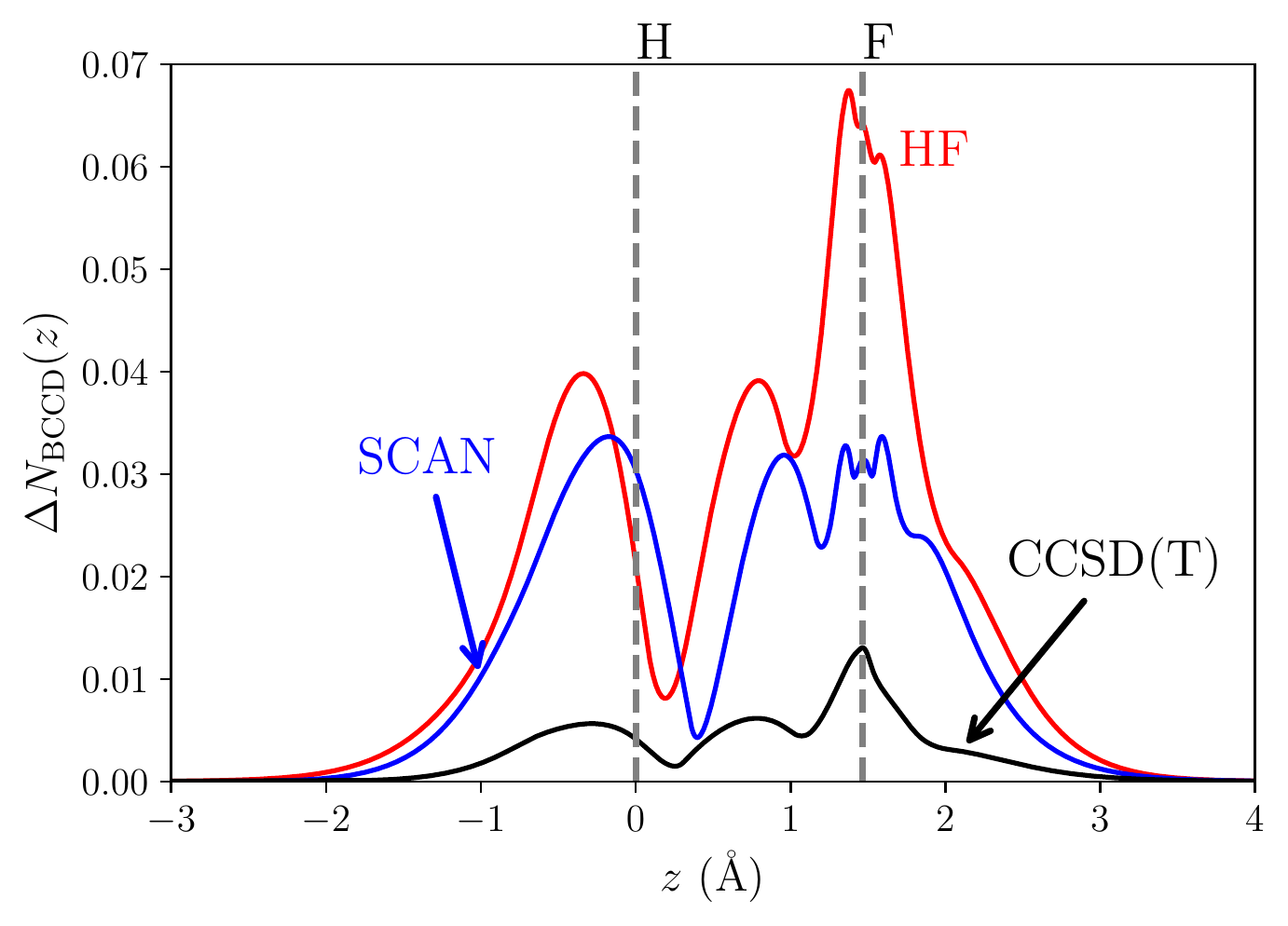}
    \caption{Integrated absolute density errors of the H...H...F transition state for Hartree-Fock (red), the SCAN meta-GGA (blue), and CCSD(T) (black). 
    Errors are evaluated with respect to the Brueckner coupled cluster doubles (BCCD) density.}
    \label{fig:bccd_cte}
\end{figure}

Instead of Hartree-Fock orbitals, a single-reference Slater determinant can be constructed from Brueckner orbitals.
In this way, one finds the determinant of maximum overlap with the true wavefunction, and eliminates the singles term in the CI expansion.
Ref. \cite{handy1989} proposed the Brueckner coupled cluster doubles (BCCD) method, which can yield more accurate electron densities than CCSD or even CCSD(T) for open-shell species like H...H...F.
Figure \ref{fig:bccd_cte} shows that, as a function of $z$ (with the $z$ axis passing through the stretched H...F bond in the transition state), the integrated absolute density error with respect to the BCCD density,
\begin{equation}
    \Delta N_\mathrm{BCCD}(z) = \int \int dx \, dy \, |n(\bm{r}) - n_\mathrm{BCCD}(\bm{r})|,
\end{equation}
is largest for the Hartree-Fock density, smaller for the SCAN density, and smaller still for the CCSD(T) density.
Note that the geometry of the transition state in Table \ref{tab:xyz_bh_RKT10} was translated and rotated so that the first hydrogen atom was located at the origin and the fluorine atom was located on the (positive) $z$ axis
After the transformation, the position of the other hydrogen atom was $(0.0, -0.36334571, -0.68412356)$ (\AA{}), and the position of the fluorine atom was $(0.0, 0.0, 1.46615748)$ (\AA{}).

%% file: tables/KS_INV_QA.tex
\begin{table}[h]
  \centering
  \begin{tabular}{lrrrrrr|rrrrrr}\hline
    & \multicolumn{6}{c|}{Forwards} & \multicolumn{6}{c}{Reverse} \\ 
    & \multicolumn{2}{c}{BH} & \multicolumn{2}{c}{DE} & \multicolumn{2}{c|}{FE} & \multicolumn{2}{c}{BH} & \multicolumn{2}{c}{DE} & \multicolumn{2}{c}{FE}\\ 
    DFA & 4Z & 5Z & 4Z & 5Z & 4Z & 5Z & 4Z & 5Z & 4Z & 5Z & 4Z & 5Z \\ \hline 
    CCSD(T) & 1.29 & 1.40 &  &  &  &  & 33.68 & 33.89 &  &  &  &  \\
 & & & & & & & & & & & & \\ 
    LSDA & -23.81 & -23.73 & -4.60 & -4.45 & -20.50 & -20.68 & 25.36 & 25.38 & -4.79 & -4.73 & -3.54 & -3.78 \\
    LSDA@HF & -5.48 & -5.00 & 13.73 & 14.28 & -20.50 & -20.68 & 43.14 & 43.19 & 12.99 & 13.08 & -3.54 & -3.78 \\
    LSDA@HF-inv & -7.45 & -7.46 & 11.76 & 11.82 & -20.50 & -20.68 & 41.15 & 41.27 & 11.00 & 11.16 & -3.54 & -3.78 \\
 & & & & & & & & & & & & \\ 
    PBE & -12.68 & -12.61 & -2.36 & -2.23 & -11.61 & -11.78 & 24.79 & 24.81 & -2.38 & -2.32 & -6.51 & -6.75 \\
    PBE@PBE-inv & -12.48 & -12.48 & -2.16 & -2.10 & -11.61 & -11.78 & 25.21 & 25.23 & -1.96 & -1.90 & -6.51 & -6.75 \\
    PBE@HF & 0.90 & 1.41 & 11.21 & 11.79 & -11.61 & -11.78 & 37.60 & 37.64 & 10.43 & 10.50 & -6.51 & -6.75 \\
    PBE@HF-inv & -1.17 & -1.18 & 9.15 & 9.20 & -11.61 & -11.78 & 35.61 & 35.66 & 8.44 & 8.52 & -6.51 & -6.75 \\
 & & & & & & & & & & & & \\ 
    SCAN & -7.48 & -7.43 & -1.08 & -0.99 & -7.70 & -7.84 & 21.94 & 22.02 & -1.35 & -1.24 & -10.39 & -10.63 \\
    SCAN@HF & 2.24 & 2.30 & 8.65 & 8.74 & -7.70 & -7.84 & 30.80 & 30.91 & 7.51 & 7.65 & -10.39 & -10.63 \\
    SCAN@HF-inv & 0.48 & 0.48 & 6.89 & 6.92 & -7.70 & -7.84 & 29.51 & 29.59 & 6.21 & 6.33 & -10.39 & -10.63 \\
 & & & & & & & & & & & & \\ 
    r$^2$SCAN & -6.97 & -6.92 & -1.09 & -1.00 & -7.18 & -7.32 & 23.72 & 23.76 & -1.34 & -1.26 & -8.63 & -8.87 \\
    r$^2$SCAN@HF & 2.50 & 2.79 & 8.38 & 8.70 & -7.18 & -7.32 & 32.56 & 32.64 & 7.51 & 7.62 & -8.63 & -8.87 \\
    r$^2$SCAN@HF-inv & 0.93 & 0.94 & 6.82 & 6.86 & -7.18 & -7.32 & 31.20 & 31.28 & 6.15 & 6.26 & -8.63 & -8.87 \\ \hline
  \end{tabular}
  \caption{
  Barrier heights (BHs), density-driven errors (DEs), and functional-driven errors (FEs) of density functional approximations (DFAs) for the forward and reverse barrier heights of the reaction H$_2$ + F $\rightarrow$ HHF $\rightarrow$ H + HF.
All units are kcal/mol.
The set of PBE and PBE@PBE-inv results measures the accuracy of our Kohn-Sham inversion of the PBE density.
Non-self-consistent ``DFA@HF'' is evaluated with the Hartree-Fock orbitals, and ``DFA@HF-inv'' is evaluated with the similar but noticeably inequivalent Kohn-Sham orbitals found by inversion of the Hartree-Fock density.
The CCSD(T) density is used as a proxy for the exact density, and has been inverted to find its Kohn-Sham orbitals.
The quadruple-$\zeta$ (4Z) and quintuple-$\zeta$ (5Z) results are obtained from orbitals (excluding those outputted by the inversion) expanded in aug-cc-pVQZ and aug-cc-pv5Z Gaussian basis sets.
Note that the density errors (kcal/mol) of the self-consistent density are much smaller in magnitude than those of the Hartree-Fock density, and of opposite sign.
    \label{tab:ks_inv_qa}
  }
\end{table}

%% file: tables/BH_devs.tex
\begin{table}[h]
  \centering
  \begin{tabular}{lrrrrr|rrrrr} \hline
    & \multicolumn{5}{c|}{Forwards} & \multicolumn{5}{c}{Reverse} \\
   Density & LSDA & PBE & BLYP & SCAN & r$^2$SCAN & LSDA & PBE & BLYP & SCAN & r$^2$SCAN \\ \hline
   Self-consistent & -25.13 & -14.01 & -12.66 & -8.83 & -8.32 & -8.51 & -9.07 & -13.52 & -11.87 & -10.12 \\
   LSDA X & -25.15 & -13.84 & -12.31 & -7.79 & -7.49 & -8.37 & -8.67 & -12.87 & -10.37 & -8.84 \\
   LSDA XC & -25.13 & -13.73 & -12.26 & -7.49 & -7.21 & -8.51 & -8.81 & -13.09 & -10.20 & -8.68 \\
   CCSD(T) & -20.68 & -11.78 & NC & -7.84 & -7.32 & -3.78 & -6.75 & NC & -10.63 & -8.87 \\
   Hartree & -20.25 & -9.34 & -5.97 & -3.18 & -2.96 & -12.83 & -12.59 & -15.06 & -14.03 & -12.75 \\
   HF & -6.82 & -0.48 & 0.67 & 0.66 & 1.15 & 9.30 & 3.75 & -0.99 & -2.98 & -1.25 \\ \hline
  \end{tabular}
  \caption{ Deviations from the CCSD(T)/aug-cc-pV5Z barrier heights (BHs): 1.40 kcal/mol for the forward BH, and 33.89 kcal/mol for the reverse BH.
  All results are computed with the aug-cc-pV5Z basis set in PySCF.
  The column indicates the density functional approximation (DFA) used to compute total energies.
  The row indicates the density used as input to the DFA energy functional.
  Thus if column = DFA1 and row = DFA2, the value in their intersection is BH(DFA1@DFA2).
  LSDA X indicates only the exchange part of LSDA \cite{kohn1965} was used; LSDA XC uses the parameterization of Ref. \cite{perdew1992} for the LSDA correlation energy.
  The Hartree approximation uses precisely zero exchange-correlation energy.
  ``NC'' indicates a value was not computed.
  \label{tab:BH_devs}
  }
\end{table}

%% file: tables/toten.tex
\begin{longtable}{llrrrrr}
  \caption{
  Total energies, in Hartree atomic units (1 hartree = 27.21139 eV = 627.509 kcal/mol), for the systems used in the calculation of the reaction barrier height H$_2$ + F $\rightarrow$ HHF $\rightarrow$ H + HF.
  All calculations employ the aug-cc-pV5Z basis set \cite{dunning1989}, and were performed in PySCF \cite{sun2020}, except for the DFA@CCSD(T) calculations, which used the code of Refs. \cite{kanungo2019,kanungo2021,kanungo2023}.  \label{tab:toten}
  }
\\ \hline 
DFA & Density & H$_2$ & F & H...H...F & H & HF \\ \hline 
\endfirsthead
\\ \hline 
DFA & Density & H$_2$ & F & H...H...F & H & HF \\ \hline 
\endhead
HF & HF & -1.13359949 & -99.41608754 & -100.52863611 & -0.49999478 & -100.07064782 \\
 \\ 
LSDA & LSDA & -1.13733363 & -99.11138401 & -100.28653041 & -0.47870532 & -99.84826732 \\ 
LSDA & SCAN & -1.13682415 & -99.10642452 & -100.27889003 & -0.47806048 & -99.84397101 \\ 
LSDA & r$^2$SCAN & -1.13682416 & -99.10632779 & -100.27883659 & -0.47806048 & -99.84387341 \\ 
LSDA & r$^2$SCAN50 & -1.13671774 & -99.10250351 & -100.26484726 & -0.47792381 & -99.83947580 \\ 
LSDA & CCSD(T) & -1.13663476 & -99.10630864 & -100.27366660 & -0.47773121 & -99.84391692 \\ 
LSDA & HF & -1.13661394 & -99.09693235 & -100.24218447 & -0.47771733 & -99.83329445 \\
 \\ 
PBE & PBE & -1.16668574 & -99.67557479 & -100.86235157 & -0.49998481 & -100.40190987 \\ 
PBE & LSDA & -1.16628304 & -99.67316627 & -100.85909791 & -0.49953525 & -100.39952431 \\ 
PBE & SCAN & -1.16654789 & -99.67427361 & -100.86025270 & -0.49985087 & -100.40077625 \\ 
PBE & r$^2$SCAN & -1.16654790 & -99.67424862 & -100.86025539 & -0.49985087 & -100.40074308 \\ 
PBE & r$^2$SCAN50 & -1.16625388 & -99.67133012 & -100.84995037 & -0.49966199 & -100.39709382 \\ 
PBE & CCSD(T) & -1.16626851 & -99.67428784 & -100.85709812 & -0.49942655 & -100.40091401 \\ 
PBE & HF & -1.16590902 & -99.66683294 & -100.83127219 & -0.49941168 & -100.39183750 \\
 \\ 
BLYP & BLYP & -1.17029653 & -99.76767888 & -100.95592035 & -0.49790802 & -100.49046511 \\ 
BLYP & LSDA & -1.16963694 & -99.76536390 & -100.95230041 & -0.49737068 & -100.48807657 \\ 
BLYP & r$^2$SCAN50 & -1.16944614 & -99.76147619 & -100.94135682 & -0.49752803 & -100.48326067 \\ 
BLYP & HF & -1.16898224 & -99.75614520 & -100.92183182 & -0.49725089 & -100.47700803 \\
 \\ 
SCAN & SCAN & -1.17189611 & -99.74828359 & -100.93201190 & -0.50016735 & -100.46692986 \\ 
SCAN & LSDA & -1.17138905 & -99.74334859 & -100.92444348 & -0.49952831 & -100.46266233 \\ 
SCAN & PBE & -1.17175679 & -99.74697060 & -100.92986129 & -0.50003230 & -100.46579158 \\ 
SCAN & r$^2$SCAN & -1.17189611 & -99.74825089 & -100.93195526 & -0.50016735 & -100.46688866 \\ 
SCAN & r$^2$SCAN50 & -1.17178240 & -99.74708333 & -100.92668628 & -0.50011939 & -100.46532954 \\ 
SCAN & CCSD(T) & -1.17181547 & -99.74808696 & -100.93016228 & -0.50000553 & -100.46721869 \\ 
SCAN & HF & -1.17152048 & -99.74417960 & -100.91241194 & -0.49999389 & -100.46167259 \\
 \\ 
r$^2$SCAN & r$^2$SCAN & -1.17189611 & -99.73323661 & -100.91616006 & -0.50016735 & -100.45385859 \\ 
r$^2$SCAN & LSDA & -1.17138905 & -99.72821708 & -100.90885686 & -0.49952831 & -100.44949812 \\ 
r$^2$SCAN & PBE & -1.17175679 & -99.73189968 & -100.91409175 & -0.50003230 & -100.45268590 \\ 
r$^2$SCAN & SCAN & -1.17189611 & -99.73320993 & -100.91610280 & -0.50016735 & -100.45381763 \\ 
r$^2$SCAN & r$^2$SCAN50 & -1.17178240 & -99.73211619 & -100.91092801 & -0.50011939 & -100.45235228 \\ 
r$^2$SCAN & CCSD(T) & -1.17181547 & -99.73304180 & -100.91428487 & -0.50000550 & -100.45414815 \\ 
r$^2$SCAN & HF & -1.17152048 & -99.72930586 & -100.89676273 & -0.49999389 & -100.44877859 \\
 \\ 
r$^2$SCAN50 & r$^2$SCAN50 & -1.17052262 & -99.71927713 & -100.88599264 & -0.50003548 & -100.42771947 \\
 \\ 
CCSD(T) & CCSD(T) & -1.17425205 & -99.70003325 & -100.87204956 & -0.49999478 & -100.42605488 \\ \hline 
\end{longtable}

%% file: tables/CBS_BH.tex
\begin{table}[h]
  \centering
  \begin{tabular}{llrrrr}\hline
& \multicolumn{2}{c}{Forward}& \multicolumn{2}{c}{Reverse} \\ 
Method & 5$\zeta$ & CBS & 5$\zeta$ & CBS \\ \hline 
HF & 13.21 & 13.22 & 26.36 & 26.36 \\ 
LSDA & -23.73 & -23.71 & 25.38 & 25.37 \\ 
PBE & -12.61 & -12.59 & 24.81 & 24.82 \\ 
SCAN & -7.42 & -7.44 & 22.02 & 22.06 \\ 
r$^2$SCAN & -6.92 & -6.93 & 23.76 & 23.79 \\ 
CCSD(T) & 1.40 & 1.46 & 33.89 & 34.09 \\ 
Ref. \cite{goerigk2017} & & 1.60 & & 33.80 \\ \hline 
  \end{tabular}
  \caption{
    Quintuple-$\zeta$ (5$\zeta$) and complete basis-set (CBS) extrapolated barrier heights for the reaction H$_2$ + F $\rightarrow$ H + HF.
    5$\zeta$ values are computed with the aug-cc-pV5Z basis set \cite{dunning1989}.
    All units are kcal/mol.
    For the CBS single-point total energies used to compute these barrier heights, see Table \ref{tab:cbs_pars}.
    \label{tab:cbs_bh}
}
\end{table}

%% file: tables/DF_pars.tex
\begin{longtable}{llrrr}
  \caption{
    Fit parameters used to extrapolate HF and DFT energies to the complete basis set (CBS) limit.
    All single-point total energies were fitted to Eq. (\ref{eq:cbs_exp}) for the values $\ell_\mathrm{max} \in \{ 4,5,6 \}$.
    To extrapolate CCSD(T) energies, we use Eq. \ref{eq:cw_extrap} and $\ell_\mathrm{max}^{(1)}=4, \, \ell_\mathrm{max}^{(2)}= 5$. 
    All units are hartree.
    \label{tab:cbs_pars}
}
 \\ \hline
Method & System & $E_\mathrm{CBS}$ & $A$ & $B$ \\ \hline 
\endfirsthead
\\ \hline
Method & System & $E_\mathrm{CBS}$ & $A$ & $B$ \\ \hline 
\endhead
HF & H$_2$ & -1.13361739 & 0.87853520 & 2.16015135 \\  
HF & F & -99.41629864 & 26.74366349 & 2.34989458 \\  
HF & H...H...F & -100.52885469 & 25.58684531 & 2.33408722 \\  
HF & H & -0.49999976 & 0.58885807 & 2.33640232 \\  
HF & HF & -100.07086679 & 24.20261675 & 2.32261158 \\ 
 \\ 
LSDA & H$_2$ & -1.13734948 & 1.08334676 & 2.22648547 \\  
LSDA & F & -99.11185509 & 6.16844607 & 1.89598763 \\  
LSDA & H...H...F & -100.28698716 & 6.85436454 & 1.92325023 \\  
LSDA & H & -0.47871028 & 0.36180145 & 2.23970304 \\  
LSDA & HF & -99.84871204 & 7.40423271 & 1.94402736 \\ 
 \\ 
PBE & H$_2$ & -1.16670250 & 0.95443643 & 2.18994285 \\  
PBE & F & -99.67603495 & 4.72331733 & 1.84728951 \\  
PBE & H...H...F & -100.86280364 & 5.20423301 & 1.87022729 \\  
PBE & H & -0.49998909 & 1.48713133 & 2.55135343 \\  
PBE & HF & -100.40236212 & 5.06031048 & 1.86453938 \\ 
 \\ 
SCAN & H$_2$ & -1.17191274 & 1.10948229 & 2.22154730 \\  
SCAN & F & -99.74861341 & 10.24790881 & 2.06881121 \\  
SCAN & H...H...F & -100.93238082 & 7.70527049 & 1.98936601 \\  
SCAN & H & -0.50017271 & 0.41753195 & 2.25273020 \\  
SCAN & HF & -100.46736190 & 5.21299281 & 1.87963189 \\ 
 \\ 
r$^2$SCAN & H$_2$ & -1.17191275 & 1.10944418 & 2.22153862 \\  
r$^2$SCAN & F & -99.73349686 & 25.71614855 & 2.30020156 \\  
r$^2$SCAN & H...H...F & -100.91644669 & 20.23179091 & 2.23291768 \\  
r$^2$SCAN & H & -0.50017271 & 0.41753262 & 2.25273407 \\  
r$^2$SCAN & HF & -100.45418081 & 14.09831613 & 2.13726669 \\ 
 \\ 
CCSD(T) & H$_2$ & -1.17452932 & & \\ 
CCSD(T) & F & -99.71465724 & & \\ 
CCSD(T) & H...H...F & -100.88686225 & & \\ 
CCSD(T) & H & -0.49999976 & & \\ 
CCSD(T) & HF & -100.44118171 & & \\ \hline
\end{longtable}

%% file: tables/H2_geom.tex
\begin{table}[h]
  \centering
  \begin{tabular}{lrrr} \hline 
    Atom & $x$ (\AA{}) & $y$ (\AA{}) & $z$ (\AA{}) \\ \hline 
    H & 0.00000000 & 0.00000000 & 0.37093843 \\ 
    H & 0.00000000 & 0.00000000 & -0.37093843 \\ \hline
  \end{tabular}
  \caption{
    The geometry of H$_2$ used for all calculations in this work.
    All geometries were taken from Ref. \cite{goerigk2017}.
    \label{tab:xyz_bh_H2}
  }
\end{table}

%% file: tables/hf_geom.tex
\begin{table}[h]
  \centering
  \begin{tabular}{lrrr} \hline 
    Atom & $x$ (\AA{}) & $y$ (\AA{}) & $z$ (\AA{}) \\ \hline 
    F & 0.00000000 & 0.00000000 & 0.45769122 \\ 
    H & 0.00000000 & 0.00000000 & -0.45769122 \\ \hline
  \end{tabular}
  \caption{
    The geometry of HF used for all calculations in this work.
    All geometries were taken from Ref. \cite{goerigk2017}.
    \label{tab:xyz_bh_hf}
  }
\end{table}

%% file: tables/RKT10_geom.tex
\begin{table}[h]
  \centering
  \begin{tabular}{lrrr} \hline 
    Atom & $x$ (\AA{}) & $y$ (\AA{}) & $z$ (\AA{}) \\ \hline 
    H & 0.14656781 & -0.24726460 & 0.00000000 \\ 
    F & 0.00000000 & 1.21154849 & 0.00000000 \\ 
    H & -0.14656781 & -0.96428389 & 0.00000000 \\ \hline
  \end{tabular}
  \caption{
    The geometry of H...H...F used for all calculations in this work.
    All geometries were taken from Ref. \cite{goerigk2017}.
    \label{tab:xyz_bh_RKT10}
  }
\end{table}

%% file: tables/H2O_geom.tex
\begin{table}[!h]
  \centering
  \begin{tabular}{lrrr} \hline 
    Atom & $x$ (\AA{}) & $y$ (\AA{}) & $z$ (\AA{}) \\ \hline 
    O & 0.0000000 & 0.0000000 & -0.3893611 \\ 
    H & 0.7629844 & 0.0000000 & 0.1946806 \\ 
    H & -0.7629844 & 0.0000000 & 0.1946806 \\ \hline
  \end{tabular}
  \caption{
    The geometry of H$_2$O used for all calculations in this work.
    All geometries were taken from Ref. \cite{rezac2008}.
    \label{tab:xyz_bh_H2O}
  }
\end{table}

%% file: tables/H2O_2_geom.tex
\begin{table}[!h]
  \centering
  \begin{tabular}{lrrr} \hline 
    Atom & $x$ (\AA{}) & $y$ (\AA{}) & $z$ (\AA{}) \\ \hline 
    O & -1.62893 & -0.04138 & 0.37137 \\ 
    H & -0.69803 & -0.09168 & 0.09337 \\ 
    H & -2.06663 & -0.73498 & -0.13663 \\ 
    O & 1.21457 & 0.03172 & -0.27623 \\ 
    H & 1.44927 & 0.91672 & -0.58573 \\ 
    H & 1.72977 & -0.08038 & 0.53387 \\ \hline
  \end{tabular}
  \caption{
    The geometry of (H$_2$O)$_2$ used for all calculations in this work.
    All geometries were taken from Ref. \cite{rezac2008}.
    \label{tab:xyz_bh_H2O_2}
  }
\end{table}